%% file: main.tex
\documentclass[12pt]{article}
\usepackage[utf8]{inputenc}
\usepackage{authblk}
\usepackage{setspace}
\usepackage[margin=1.25in]{geometry}
\usepackage{graphicx}
\usepackage{subcaption}
\usepackage{newtx,newtxmath}
\usepackage{lineno}
\usepackage{hyperref}
\hypersetup{
	colorlinks=true,
	linkcolor=blue,
	filecolor=violet,     
	urlcolor=blue,
	citecolor=red
}
\usepackage[backend=bibtex,style=nejm, 
citestyle=numeric-comp,
sorting=none]{biblatex}
\input{command}

\begin{document}
\input{content}
\end{document}

%% file: command.tex
\newcommand{\es}{\text{EoS}}

\newcommand{\nb}{\ensuremath{ \nabla }}

\newcommand{\m}{\ensuremath{{\mu \nu}}}

\newcommand{\p}{\ensuremath{\partial{}}}

\newcommand{\cs}{\ensuremath{c_{s}^2}}

%% file: content.tex
\title{Dynamical Analysis of the Dirac-Born-Infeld type of Tachyon field minimally coupled with barotropic fluid using EOS parametrization of a field 
}
\author[1*]{Saddam Hussain}
\author[2]{Saikat Chakraborty}
\author[3]{Nandan Roy}
\author[4]{Kaushik Bhattacharya}
\affil[1,4]{Department of Physics, Indian Institute of Technology, Kanpur\\ Uttar Pradesh 208016, India.}
\affil[2]{The Institute of Fundamental Study ``The Tah Poe Academia Institute'', Naresuan University, Phitsanulok 65000, Thailand}
\affil[3]{Centre for Theoretical Physics and Natural Philosophy, Mahidol University, Nakhonsawan Campus, Phayuha Khiri, Nakhonsawan 60130, Thailand}
\affil[*]{msaddam@iitk.ac.in}

\onehalfspacing
\maketitle

\date{}

\begin{abstract}

In this paper, we present a dynamical system analysis of the tachyon dark energy model by parametrization of the equation of state (EoS) of the dark energy. The choice of parametrization can constrain the form of the field potential, and as a result, the theory can be directly constrained from the observation without assuming a particular form of the potential. 

\end{abstract}

\section{Introduction: the tachyon dynamics}
\begin{figure}[t]
	\begin{minipage}{0.5\linewidth}
		\centering
		\includegraphics[scale=0.45]{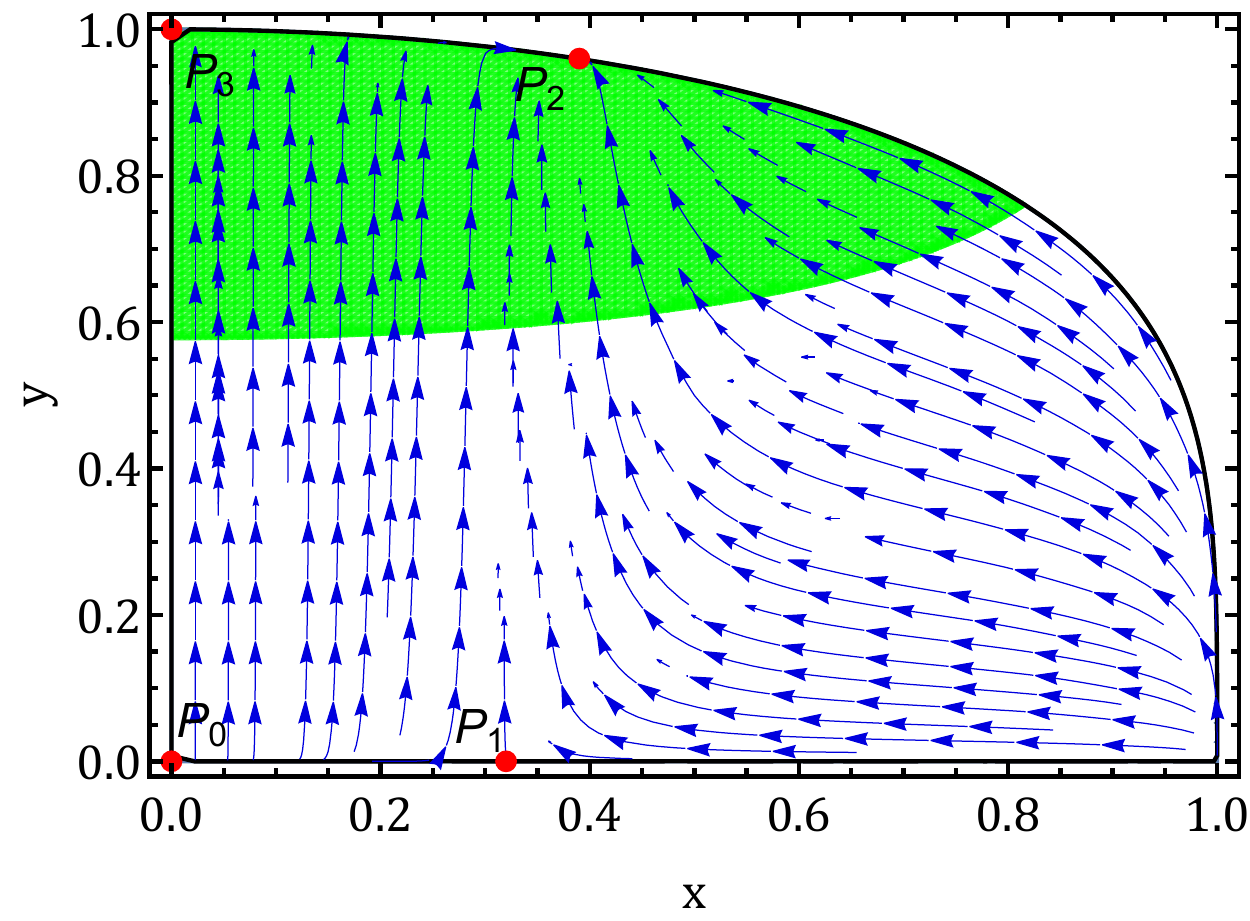}
		\caption{The phase space plot for $ \omega_{0} =-0.9	$ and $\omega_{1} = - 0.05$, where green region shows the accelerating region $-1 \le \omega_{\rm tot}=-1- \frac{2\dot{H}}{3H^2}<-1/3$. }
		\label{fig: 3rd_para_phase_space_eps_1}
	\end{minipage}
	\hspace{0.3cm}
	\begin{minipage}{0.5\linewidth}
		\centering
		\includegraphics[scale=0.45]{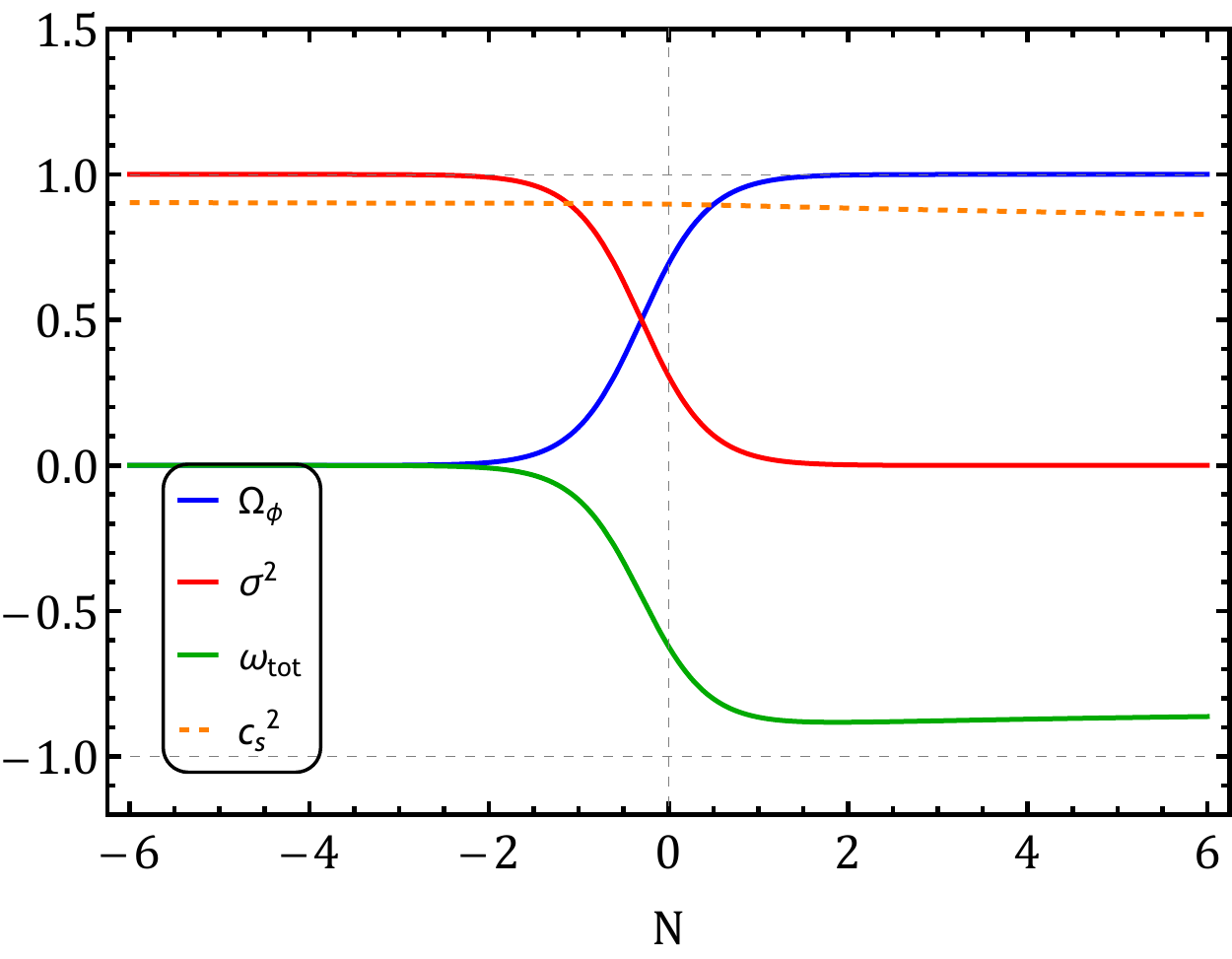}
		\caption{The evolution plot for $\omega_{0} =-0.9	$ and $\omega_{1} = - 0.05$ where field density $\Omega_{\phi} = y^2(1-x^2)^{-1/2}$, fluid density $\sigma^2 = 1 - \Omega_{\phi}$ and sound speed $0 \le \cs = \delta P_{\phi}/\delta \rho_{\phi} \le 1$.  }
		\label{fig: 3rd_para_evo_eps_1}
	\end{minipage}
	
\end{figure}

The action of the field was first discovered in string theory\cite{Sen:2002in,Sen:2002nu,Sen:2002an} and later on it has been modeled to study the early time epoch as inflaton field\cite{delCampo:2009ma,Herrera:2006ck,Mohammadi:2018oku,Sadeghi:2008wy,Abramo:2003cp} and a dark energy (DE) candidate in the late-time epoch\cite{Mohammadi:2020vgs,Calcagni:2006ge,Sheykhi:2011cn,Panda:2005sg,Shao:2007zv,Teixeira:2019tfi,Avelino:2011ey,Copeland:2004hq}. The action of the tachyon field is a Dirac-Born-Infeld (DBI) type and dark matter is characterized as pressureless perfect fluid; both are minimally coupled with gravity as follows: 
\begin{equation}
	S = 	\int_\Omega d^4x \sqrt{-g} \left[\frac{R}{2\kappa^2}  -V(\phi) \sqrt{ 1 +(\nb_{\mu}\phi) (\nb^{\mu}\phi)}  \right] + S_{M}
	\label{tachyon_action}
\end{equation}
where the first term in the action is the Einstein-Hilbert action, the second term corresponds to tachyon field, and the final term is the perfect fluid action\footnote{$\kappa^2 = 8\pi G$ is the inverse of reduced Planck mass.}. We conduct the analysis in the FLRW  background metric $ds^2 = -dt^2 + a(t)^2 d\vec{x}^2$, where \(a \) is the scale factor; this is consistent with the observational fact that the universe appears flat, homogeneous, and isotropic at the largest scale. Due to the highly symmetric metric the field $\phi$ only depends on time. To analyze the dynamics of the action, it is necessary to obtain the equation of motion by varying the action with respect to the metric and field $(\delta g^{\m}, \delta\phi)$. Variation of action with respect to metric yields Einstein field equation, i.e., $G_{\m} = \kappa^2 (T_{\m}^{M} + T_{\m}^{\phi})$ and field equation of motion is $	\ddot{\phi} + 3 H \dot{\phi} (1 -  \dot{\phi}^2) +  \frac{V_{,\phi}}{V} (1 - \dot{\phi}^2) = 0.$\footnote{$\dot{()}\equiv d()/dt$, $V_{,\phi}\equiv \p V/\p \phi$  and $H = \dot{a}/a$ is a Hubble parameter.} The energy density of the field is $\rho_{\phi} ={V(\phi)}({1 -  \dot{\phi}^2})^{-1/2}$, and pressure is $P_{\phi}= - V(\phi) \ ({1 -  \dot{\phi}^2})^{1/2} $ and 
the field is to be depicted as dark energy it should exert negative pressure for which $V(\phi)>0$ and $\dot{\phi}^2\ll 1$. Typically, the acceleration is denoted by the \es\ which is defined as the ratio of pressure to energy density. In the case of the tachyon field, the \es\ is $\omega_{\phi} = -1 + \dot{\phi}^2$. Since, the field is a dynamical quantity, its potential plays a crucial role in achieving late-time cosmic acceleration provided $\dot{\phi}^2 \ll 1$. In this scenario, a broad class of potentials could be investigated, but doing so would not be an efficient task. Instead of choosing any specific  potential we choose some approximate form of the \es\ of the tachyon field. These equations are time dependent and parameterized phenomenologically. Such parameterized \es\ of the DE 
can circumvent the selection of the potential, enabling the system to be constrained directly by the observation\cite{Hussain:2022dhp}. In this paper, we will evaluate the dynamics of the system using the parametrization $\omega_{\rm de} = \omega_{0} + \omega_{1} (t \dot{H} / H)$,\footnote{$(\omega_{0},\omega_{1})$ are the constants.} \cite{Usmani:2008ce}. 
To accomplish this, we will first define the minimal dimensionless dynamical variables required to transform the field equation of motions into three first order differential equation known as autonomous equations. The dimensionless variables are $	x = \dot{\phi}, \ y = \frac{\kappa\sqrt{V(\phi)}}{\sqrt{3} H}, \ \lambda = - \frac{V_{,\phi}}{\kappa V^{\frac{3}{2}}}, \ \Gamma = V \frac{V_{,\phi \phi}}{V_{,\phi} ^2},\; \sigma^2 = \frac{\kappa^2 \rho_M}{3 H^2}.$ The autonomous equations are $x'=f(x,y,\lambda), y'={h}(x,y,\lambda), \lambda' = - \ \sqrt[]{3} \ \lambda^2 x y \left(\Gamma - \frac{3}{2} \right)$.\footnote{$()'$ is $d()/dN$, where $dN =d \log a$ is number of e-folds and see ref.\cite{Hussain:2022dhp} for more detail.} Therefore, to close the system, at least three variables are required, resulting in 3D phase space. $\Gamma$ depends on the choice of potential, in some instances, it can be constant or $\phi$ dependent. As, $\omega_{\rm de}$ equal to $\omega_{\phi}$, and by differentiating both side with respect to time, one obtains new $x'=f_{1}(x,y, \lambda)$. Equating $f_{1}$ with $f$, $\lambda$ in $(x,y)$ can be determined. Using this $\lambda$ in the old \(x',y'\), closes the system in only these two variables without requiring any additional variables, thereby reducing the phase space from 3D to 2D. Additionally, $\Gamma$ in $(x,y)$ can be obtained by taking the derivative of $\lambda$ and equating with its unconstrained $\lambda'$. The dynamical equations in \((x',y')\) produce two stable critical points, $P_{2}$ and $P_{3}$ shown in the phase space $(x,y)$ in Fig.[\ref{fig: 3rd_para_phase_space_eps_1}]. Point $P_{3}$ is independent of the model parameters and represents the de-Sitter phase of acceleration, whereas $P_{2}$ is a model parameter dependent $(\omega_{0},\omega_{1})$. The vector trajectories from the kinetic dominating region $\dot{\phi}^2 \sim 1$, initially attract to the the matter phase point $P_{1}$, then to $P_{2}$. All field parameters vanish at point $P_{0}$, yielding a trivial non-accelerating fixed point. In Fig.[\ref{fig: 3rd_para_evo_eps_1}], we solve the autonomous system from past to asymptotic future to discover observable parameter dynamics against $N$. In the past, the fluid density dominated field density and the effective \es\ $\omega_{\rm tot}\sim 0$, identify as the matter phase. Field density dominates and the effective \es\ approaches $-1$ as the system evolves, nevertheless the sound speed $\cs$ remain in causal limit.

\section{Conclusion}
The given parametrization is compatible with the tachyon field and produces a stable accelerating fixed point. In the presence of pressureless background fluid, the field resembles the characteristics of dark energy. In addition, the method has been examined when both the dark sectors sector interacts $\nb_{\mu}T^{\m}_{\phi} = -Q^{\nu};\nb_{\mu}T^{\m}_{M} =Q^{\nu}$ and produces the stable accelerating solution see in ref.\cite{Hussain:2022dhp}. 

\printbibliography